\documentclass[english,preprint,showpacs]{revtex4}
\usepackage{babel}
\usepackage[ansinew]{inputenc}
\usepackage{graphicx}
\usepackage{color}
\usepackage{amsfonts}
\usepackage{amsmath,amssymb}
\usepackage{amsthm,subfigure}
\usepackage{amsbsy}
\usepackage[mathscr]{eucal}

\begin{document}
\title{Estimation of length scale of RS II-$p$ braneworld model through perturbations in Helium's atom ground state energy}

\author{Nephtal\'i Garrido}\email{falgargon@gmail.com}
\author{H\'ector H. Hern\'andez}\email{hhernandez@uach.mx}
\affiliation{Universidad Aut\'onoma de Chihuahua, Facultad de
Ingenier\'\i a, Nuevo Campus Universitario, Chihuahua, Chih.,
M\'exico}

\begin{abstract}
We put to the test an effective three-dimensional electrostatic potential, obtained effectively by considering an electrostatic source inside a (5+$p$)-dimensional braneworld scenario with $p$ compact and one infinite spacial extra dimensions in the RS II-$p$ model, for $p=1$ and $p=2$. This potential is regular at the source and matches the standard Coulomb potential outside a neighborhood. We use variational and perturbative approximation
 methods to calculate corrections to the ground energy of the Helium atom
modified by this potential, by making use of a 6 and 39-parameter trial wave function of Hylleraas type for the ground state. These corrections to the ground-state energy are compared with experimental data for Helium atom in order to set bounds for the extra dimensions length scale. We find that these bounds are less restrictive than the ones obtained by Morales et. al. through a calculation using the Lamb shift in Hydrogen.
\end{abstract}


\maketitle

\section{Introduction} \label{sec: intro}

Theories that support the existence of spatial dimensions, additional to the three we observe in everyday physics,
have been proposed since the 20's decade of the last century with the attempts of Theodor Kaluza (1921) and Oscar Klein (1926) to provide a unified theory of gravity and electromagnetism \cite{kaluza, klein}. More recently, the idea of a ``braneworld'' scenario \cite{Akama, Rubakov, Visser}, where our three-dimensional world is a subspace -or ``brane''- embedded into a higher-dimensional universe known as the ``bulk'' has been widely spread, specially given the philosophy and machinery of string theory, which can only be correctly formulated in a space-time of at  least ten dimensions.

Among brane world models, there are some that consider compact and large extra dimensions \cite{Arkani}, and others comprising infinite and warped ones \cite{Randall1,Randall2}. These models have opened a gate to interesting possible solutions to long-standing problems in physics, such as the hierarchy problem and the cosmological constant problem in high-energy physics \cite{Rubakov:2001kp, intro_bw}. At the same time they allow the testing of models at the cosmological level \cite{roy}. Moreover, the search for testable, low energy effects of extra dimensions is an area of research that has been extremely active in the last years., 

With regard to the last point above, there have been several studies that explore the possibility for obtaining measurable evidence from models with extra dimensions by studying low energy phenomena. In particular we mention the ones that have been performed in the RS II-$p$ setup, such as the electric charge conservation \cite{Dubovsky:2000av}, the Casimir effect between parallel plates \cite{linares:2010, frank:2008}, the Hydrogen Lamb shift \cite{morales:2006}, the Cavendish experiment and the scattering process of electrons by Helium atoms \cite{morales:2011}. 

One of the most successful applications of the quantum-mechanical
variational and perturbative methods is the associated with the study of the Helium atom, in the non-relativistic and relativistic \footnote{The exact 
relativistic formulation for the two electron system cannot be written in closed form, and certain methods in the form of relativistic
corrections to the non-relativistic formulation as a series expansion in powers of the fine structure constant $\alpha$
 have to be applied. One of the most prominent corrections is expressed in the Breit equation (see \cite{bethe}).} formulation.
 A recent study of the corrections to the ground-state energy of the Helium atom by the presence of extra dimensions
 was performed in \cite{liu:2007}, where the ADD model was employed. In the light of such studies it results interesting to perform a similar analysis
 for the Helium atom in the RS II-$p$ model.

This work is organized as follows. In section~\ref{sec: 6d} we consider an hybrid braneworld scenario that contains both compact and infinite extra dimensions. It consits of a single (3+$p$)-brane embedded in a (3+$p$+1)-dimensional space, with $p$ compact extra dimensions and one infinite and warped, making the whole model (5+$p$)-dimensional. This is a modification to the Randal-Sundrum II scenario \cite{Randall2}, which is known in the literature as the RSII-$p$ model. The importance of this model resides in its property of localization of fields in the brane, scalar, gauge and gravitational, due to the gravitational field produced by the brane itself. Particularly, gauge fields are localized only when $p \geq 1$ \cite{Dubovsky:2000av}. Recently, it has been found that an electromagnetic source lying on the brane of the RSII-$p$ model along the compact dimensions, that looks point-like to an observer in the usual 3D subspace, produce a static potential that is not singular from the 3D perspective, and matches Coulomb's potential outside a small neighborhood \cite{morales:2006, morales:2011}. We use this modified potential as the interacting force between the electrons and the nucleus of the Helium atom, in order to obtain bounds to the characteristic length of the model.

In section~\ref{sec:energy} we obtain a 6-parameter trial wave function of the Hylleraass type \cite{hyleras}, and employ the 39-parameter function of \cite{kinoshita} . Then we use perturbations to the Coulomb potential obtained by Morales et al for the RS II-$p$ scenario \cite{morales:2006} in order to obtain corrections to the ground-state energy of the Helium atom. These corrections are given in terms of $\epsilon$, the anti-de Sitter radius of the model, and by comparing the obtained values with the experimental uncertainty of the Helium's ionization potential we compute restrictions to the value of $\epsilon$ under this kind of experiments.

Finally in section~\ref{sec:conclusions} we discuss our results, give some conclusions and the possibility of future work.

\section{(5+$p$) model and modified potentials} \label{sec: 6d}
It has been shown recently \cite{morales:2006, morales:2011} that, when one considers an static source on a (3+$p)$ brane
in the form of a $p$-dimensional torus in a Randall-Sundrum II-$p$ scenario, an observer living on the
usual 3D subspace sees, effectively, a non singular electrostatic potential that coincides with the Coulomb
potential outside a small neighborhood. This 3D potential depends on the AdS curvature radius $\epsilon$.

One can test the physical implications of this modified potential in a phenomenological way, much in the same
spirit as in \cite{morales:2006, morales:2011}, where theoretical bounds were obtained for the radius of the extra dimensions
by considering the Hydrogen Lamb shift, the Cavendish experiment for electromagnetism, and the scattering of electrons by Helium atoms. Similar 
phenomenological considerations have been put forward in a different scenario in \cite{liu:2007}: the ground state energy
for the Helium atom and Helium-like ions was computed considering the gravitational correction to the 
Arkani-Hamed-Dimopoulos-Dvali (ADD) model.

In this work we use the modified potential from the RS II-$p$ model and obtain upper bounds to the AdS radius by 
computing the ground state of the non relativistic Helium atom using a combination of the Ritz-variational and 
perturbation approximation methods.

\subsection{RS II-$p$ model and potential}

This section follows from \cite{morales:2006, morales:2011}. The Randall-Sundrum II-$p$ model consists of a (3+$p$) brane with $p$ compact dimensions in a (5+$p$) space-time. The metric of this latter space-time is comprised by two patches of AdS 5+$p$-dimensional
of Gaussian curvature $\epsilon$
\begin{equation} \label{ads metric}
 ds_{5+p} = e^{-2 |y|/ \epsilon}  \left( \eta_{\mu \nu} dx^{\mu} dx^{\nu} - \sum_{i=1}^{p} R_i^2 d\theta_i^2 \right) -dy^2
 \end{equation}
 where $\eta_{\mu \nu}$ is the 4D Minkowski metric, $\theta_i \in [0,2\pi ]$ are compact coordinates and $R_i$ the 
 compact dimensions.
 
 The regularized electrostatic potential on the 3D subspace of the 3+$p$ brane, coming from a scalar source field
 localized on the brane, is
 \begin{equation} \label{potential}
\phi(r) = \left\{
\begin{array}{rl}
\frac{2e_1}{ \pi \epsilon} \left[  \frac{\arctan{x}}{x}+ \frac {1}{1+x^2} \right], &  p=1,\\
\frac{e_2}{ \epsilon} \left[ \frac{1}{\sqrt{1+x^2} } + \frac{1}{2 \left( 1+x^2 \right)^{3/2}}  \right], & p=2
\end{array} \right.
\end{equation}
 where $x=\frac{r}{\epsilon}$, $e_1=\frac{e^{(6)}}{2 R \epsilon^2}$, $e_2= \frac{e^{(7)}}{ R^2 \epsilon^2}$, and 
 $e^{(i)}$ is the total charge in each case. $r$ is the radial coordinate in spherical coordinates with origin at
 the source.
 
 Using the limit $r \gg \epsilon$ in (\ref{potential}) one obtains
 \begin{equation} \label{coulomb}
\phi(r) = \left\{
\begin{array}{rl}
\frac{2e_1}{ \pi \epsilon} \left[  \frac{\pi}{2x}- \frac {2}{3x^4} \right], &  p=1,\\
\frac{e_2}{ \epsilon} \left[ \frac{1}{x } - \frac{3}{8 x^5}  \right], & p=2
\end{array} \right.
\end{equation}
which clearly reduces to the Coulomb potential for large $r$. It is worth noticing that these modified potentials are finite
at the source.

Now we intend to test the model by analyzing the non relativistic Helium atom ground energy.

\section{Modified ground state energy of Helium atom} \label{sec:energy}
 
 \subsection{Ritz-variational method}
The Schr\"odinger equation for the Helium atom, in atomic units, is
 \begin{equation} \label{schroedinger}
  \left[ \frac{1}{2} \left( \nabla_1^2 + \nabla_2^2 \right) +E + (\frac{1}{r_1}+ \frac{1}{r_2}- \frac{1}{r_{12}}) \right] \psi =0,
 \end{equation}
 neglecting the dynamics of the nucleus, where $r_i$ is the radial distance from the nucleus to each of the two
 electrons, and $r_{12}$ is the distance between them. Now, following \cite{hyleras, kinoshita}, it is customary to use the
 two elliptical coordinates and the interparticle distance in (\ref{schroedinger})
 \begin{equation} \label{coords}
  s=r_1+r_2, \quad t=r_1-r_2, u=r_{12}.
 \end{equation}
 
 One can integrate the angular dependence in (\ref{schroedinger}) and the volume element un these coordinates is
 \begin{equation} \label{volume}
  dV= 2\pi^2 (s^2-t^2) u ds \ dt \ du, \quad 0 \leq t \leq u \leq s < \infty.
 \end{equation}
 
 We employ a combination of the Ritz-variational and perturbative method in order to obtain the ionization
 potential for the ground state of the Helium atom.  The variational approximation for the upper bound
 of the ground state energy of a quantum mechanical system is obtained from minimizing the following integral 
 \begin{equation} \label{variational}
  E[U] = \frac{\int \psi^* H \psi \ dV}{\int |\psi|^2 \ dV}
 \end{equation}
 where $\psi$ is a suitable wave function. As a first step, and following \cite{hyleras} we propose a test function of
 the Hylleraas type with the modifications proposed by \cite{kinoshita}
 \begin{equation} \label{wavefunction}
 \psi (s, t, u) = e^{-(1/2) k s} \sum_{l,m,n = 0}^{N} C_{l, m, n} (k s)^l(k t)^{2 m} (k u)^n
 \end{equation}
 where $N$ is the number of terms in the sum and the ``effective nuclear charge" $k$ and the parameters $C_{l,m,n}$ are to be fixed by minimizing (\ref{variational}).
 
 \subsubsection{Six-parameter wave function}
 As a first attempt, we propose a six-parameter trial function. Equation (\ref{variational}), using (\ref{volume}) and (\ref{coords}) 
 translates into
 \begin {equation} \label{integral}
 E= \frac{k^2 M - kL}{N},
 \end{equation}
 with
 \begin{eqnarray} \label{LMN}
 L &=& \int_0^{\infty} \int_0^s \int_0^u ds\ du \ dt\ (4su-s^2+t^2) \psi^2, \nonumber \\ 
 M &=& \int_0^{\infty} \int_0^s \int_0^u ds\ du \ dt\ \left\{ u(s^2-t^2) \left( \nabla \psi \right)^2 + 
  2s(u^2-t^2) \psi_s \psi_u +2t (s^2-t^2) \psi_t \psi_u \right\},  \nonumber \\
  N &=& \int_0^{\infty} \int_0^s \int_0^u ds\ du \ dt\ u (s^2-t^2) \psi^2.
 \end{eqnarray}
 
 The resulting values of the parameters are shown in Table  \ref{tabla1}
 
 \begin{table}[htbp] 
 \label{tabla1}
 \begin{center}
 \begin{tabular}[c]{|c|c|}
 \hline
 Parameter & Value \\
 \hline
 \hline
 $k$ & 3.7211124 \\
 \hline
 $10\, C_{0,0,1}$ & 0.972\\
 \hline
 $10\, C_{1,0,0}$ & -0.277\\
 \hline
 $100\, C_{0,1,0}$ & 0.97\\
 \hline
 $100\, C_{0,0,2}$ & -0.24\\
 \hline
 $100\, C_{2,0,0}$ & 0.25\\
 \hline
 \end{tabular}
 \end{center}
 \caption{{\small Values of the six parameters for the trial wave function of the ground state of Helium. }}
\end{table}
We also employ the 39 parameters wave function obtained by Kinoshita \cite{kinoshita}, which in turn provides one of the most accurate results available in the estimation of the ground energy.
 
 \subsection{Perturbative approximation for the modified potential}
 We now perform the perturbative method to first order to compute the correction to the ground energy
 of He. The modified Hamiltonian can be written in the form
 \begin{equation} \label{perturbation}
  H= H_0 +H',
 \end{equation}
 where $H_0$ is the original Hamiltonian (\ref{schroedinger}), while $H'$ comes from (\ref{coulomb})

 \begin{eqnarray} \label{perturbed hamilton}
 H_5^{'} &=& -\frac{4e \epsilon^3}{3\pi } \left( \frac{32}{(s-t)^4}+\frac{32}{(s+t)^4}-\frac{1}{u^4} \right) , \nonumber \\
 H_6^{'} &=& -\frac{3e \epsilon^4}{8} \left( \frac{64}{(s-t)^5}+\frac{64}{(s+t)^5}-\frac{1}{u^5} \right),
 \end{eqnarray}
 where we used (\ref{coords}).
 
 From first-order perturbation theory we know that the expected value of the Hamiltonian in (\ref{perturbation}) 
 is the correction to the energy of the ground state
\begin{equation}\label{delta H}
\Delta E = \int \psi^*H'\psi.
\end{equation}

We use the trial wave function obtained in the previous section, Eq.(\ref{wavefunction}), and Kinoshita's 39-parameter function
obtained in \cite{kinoshita} to evaluate the correction (\ref{delta H}) and, in turn, obtain the estimations for the radius of the
extra dimensions $\epsilon$, and then contrasting with the uncertainty of the most accurate experimental value for the Helium's ionization potential  \cite{kinoshita}.
 The results are given in Table \ref{tabla2}.
\begin{table}[htbp]  
\label{tabla2}
 \begin{center}
 \begin{tabular}[c]{|c|c|c|}
 \hline
  & $p$=1 & $p$=2 \\
 \hline
 $\psi_6$ & $\epsilon \leq 5.9 \times 10^{-5}$ & $\epsilon \leq 1.3 \times 10^{-4}$\\
 \hline
 $\psi_{39}$ & $\epsilon  \leq 2 \times 10^{-5}$ &  $\epsilon \leq 9.6 \times 10^{-5}$\\
 \hline \end{tabular}
 \end{center}
 \caption{{\small Estimated values (in meters) for the extra dimensional radius $\epsilon$ from perturbation method 
 to first order. }}
\end{table}

These values for the extra dimensions radius are less stringent than those obtained in \cite{morales:2006} and \cite{morales:2011}.
 
 \section{Conclusions} \label{sec:conclusions}
In this work we have obtained estimations for the radius of the extra dimensions of a Randall-Sundrum II-$p$ model,
for $p=1,2$, by computing the perturbed modifications to the ground-state energy of the Helium atom considering
the regularized effective electromagnetic potential, as seen by a 3D observer living on the brane. We used, as the
trial function for the perturbation, a wave function of the Hylleraas type with six parameters and the one obtained
in \cite{kinoshita} for the ground state with 39 parameters.

For the case of the six-parameter function the most accurate result is $\epsilon \sim 10^{-5}$, which is very similar to
the one obtained for the 39-parameter function. One can see that the use of a wave function with more parameters 
 gives a more accurate result for the ground energy, as has been pointed out by \cite{kinoshita}, although in the present work this does
 not represent a necessarily better bound for the radius of the extra dimensions.
 
 As for the result itself, ours is less stringent than the one obtained in the original work \cite{morales:2006} and 
 \cite{morales:2011}, where values reported for the upper bound of $\epsilon$ range from $10^{-14} $m  to $10^{-7}$ m.
  
There are some possibilities to explore in order to close the gap between these theoretical values: extend the present work by including the polarization mass terms, and relativistic corrections,  which by itself is a very interesting analysis, though rather involved. It is known that the inclusion of such corrections in general add to the precision of the uncertainty, and may provide a more  restricted size for the extra dimensions. Work in this direction is currently under way.
 
It is also interesting to extend our analysis to more general scenarios, like the studio of ionization energies for several
light ($\textrm{Li}^+, \textrm{Be}^+$, etc.) ions, the determination of ionization energy in the interaction 
between molecules formed by atoms, among others. All these systems provide excellent grounds for testing
the predictions made by higher dimensional theories because experimental values are well measured and 
one can actually perform laboratory experiments in order to contrast experimental and theoretical data.

\section*{Acknowledgments}
This work was supported by Mexico's National Council for Science and Technology (CONACyT) grant CONACyT-CB
2008-01-101774.

\end{document}